\documentclass[sigconf,anonymous=false]{acmart}

\AtBeginDocument{%
  }

\copyrightyear{2026}
\acmYear{2026}
\setcopyright{cc}
\setcctype{by}
\acmConference[SIGIR '26]{Proceedings of the 49th International ACM SIGIR Conference on Research and Development in Information Retrieval}{July 20--24, 2026}{Melbourne, VIC, Australia}
\acmBooktitle{Proceedings of the 49th International ACM SIGIR Conference on Research and Development in Information Retrieval (SIGIR '26), July 20--24, 2026, Melbourne, VIC, Australia}
\acmDOI{10.1145/3805712.3808619}
\acmISBN{979-8-4007-2599-9/2026/07}

\usepackage[show]{chato-notes}
\usepackage{tcolorbox}
\usepackage{listings}
\tcbuselibrary{listings,breakable}
\usepackage{multirow}
\usepackage{svg}
\usepackage{amsmath}
\usepackage{bbm}
\usepackage{marginnote}
\usepackage[x11names]{xcolor}

\newcommand{\jm}[1]{\textcolor{black}{#1}}
\newcommand{\crc}[1]{\textcolor{black}{#1}}

\usepackage{marginnote}

\newcommand{\githublink}[2]{%
  \href{#1}{%
    \includegraphics[width=1em,height=1em]{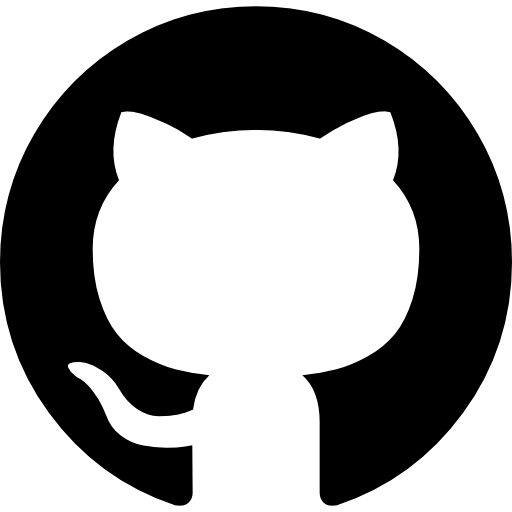}\,
    \texttt{#2}%
  }%
}

\newcommand{\hflink}[2]{%
  \href{#1}{%
    \includegraphics[width=1em,height=1em]{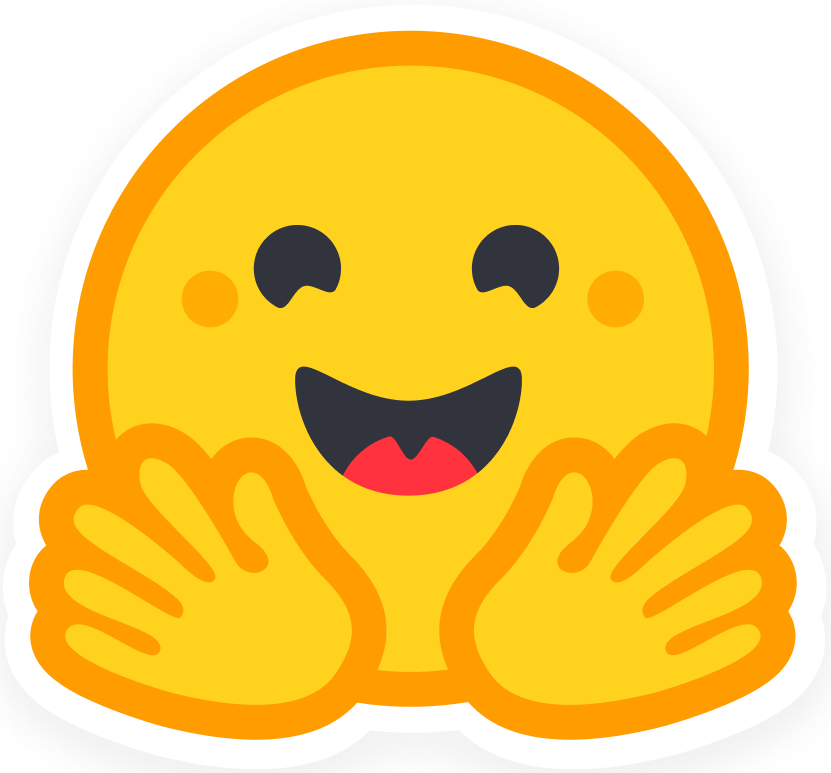}\,
    \texttt{#2}%
  }%
}

\newcommand{\observation}[2]{%
\begin{tcolorbox}[
  width=\linewidth,
  colback=blue!5!white,
  colframe=blue!75!black,
  left=0pt,
  right=0pt,
  top=0pt,
  bottom=0pt,
  boxrule=0.8pt,
  title=#1
]
#2
\end{tcolorbox}
}

\newcommand{\nonsensitivedoc}[2]{%
\begin{tcolorbox}[
  width=\linewidth,
  colback=green!5!white,
  colframe=green!75!black,
  left=0pt,
  right=0pt,
  top=0pt,
  bottom=0pt,
  boxrule=0.8pt,
  title=#1
]
#2
\end{tcolorbox}
}

\newcommand{\sensitivedoc}[2]{%
\begin{tcolorbox}[
  width=\linewidth,
  colback=red!5!white,
  colframe=red!75!black,
  left=0pt,
  right=0pt,
  top=0pt,
  bottom=0pt,
  boxrule=0.8pt,
  title=#1
]
#2
\end{tcolorbox}
}

\lstdefinelanguage{python}{
  keywords={import,from,as,def,return,class,for,while,if,elif,else,with},
  keywordstyle=\color{blue},
  commentstyle=\color{gray},
  stringstyle=\color{orange},
  sensitive=true
}

\newtcblisting{pycode}{
  listing engine=listings,
  listing only,
  breakable,
  colback=gray!5,
  colframe=black,
  boxrule=0.5pt,
  arc=2mm,
  left=6pt,
  right=6pt,
  top=6pt,
  bottom=6pt,
  listing options={
    language=python,
    basicstyle=\ttfamily\small,
    numbers=left,
    numberstyle=\tiny,
    stepnumber=1,
    numbersep=5pt,
    showstringspaces=false,
    breaklines=true
  }
}

\lstdefinestyle{mystyle}{
    language=Python,
    basicstyle=\ttfamily\footnotesize,
    keywordstyle=\color{blue}\bfseries,
    commentstyle=\color{gray}\itshape,
    stringstyle=\color{red},
    numberstyle=\tiny\color{gray},
    numbers=left,
    stepnumber=1,
    numbersep=5pt,
    backgroundcolor=\color{lightgray!20},
    frame=single,
    rulecolor=\color{black},
    breaklines=true,
    breakatwhitespace=true,
    captionpos=b,
    tabsize=4,
    showspaces=false,
    xleftmargin=10pt,    
    aboveskip=5pt,      
    belowskip=5pt,      
    columns=flexible,    
}
\begin{document}

\title[A Sensitivity-Aware Test Collection for Search Among Personal Information]{A Sensitivity-Aware Test Collection for \\Search Among Personal Information}

\author{Jack McKechnie}
\email{j.mckechnie.1@research.gla.ac.uk}
\affiliation{%
  \institution{University of Glasgow}
  \city{Glasgow}
  \country{Scotland}
}

\author{Graham McDonald}
\email{graham.mcdonald@glasgow.ac.uk}
\affiliation{%
  \institution{University of Glasgow}
  \city{Glasgow}
  \country{Scotland}
}

\author{Craig Macdonald}
\email{craig.macdonald@glasgow.ac.uk}
\affiliation{%
  \institution{University of Glasgow}
  \city{Glasgow}
  \country{Scotland}
}

\renewcommand{\shortauthors}{Jack McKechnie, Graham McDonald, \& Craig Macdonald}

\begin{abstract}
\looseness -1
Traditional search tasks aim to satisfy user information needs by returning a subset of a collection of documents, ranked by the documents' relevance to a user query. However, some collections that contain useful information also contain \emph{sensitive} personal information. Recently, there has been increasing interest in the development of \emph{Sensitivity-Aware Search} (SAS) retrieval models to provide users with effective retrieval results without revealing such sensitive information. To develop such systems, test collections containing both sensitive and non-sensitive information, a set of queries, and query-document relevance assessments are required. The Enron email corpus contains real business-related emails, where some emails also contain sensitive personal information. However, the original Enron collection does not contain queries or query-relevance assessments. To this end, we crowdsource 150 query formulations for 50 different topics and 11,471 query-relevance assessments for a subset of the Enron documents that have been manually labelled for sensitivity. \jm{We follow best practices for using large language models (LLMs) in Information Retrieval evaluation to extend the collection further with additional LLM judged query-relevance assessments and sensitivity labels. We present baseline performances for relevance, sensitivity classification, and sensitivity-aware search on the collection.} We make the collection available, including through the popular ir\_datasets package, and provide pre-built sparse and dense indices on Huggingface to facilitate easy experimentation.

\phantom{1}

{\centering
\includegraphics[width=1.25em,height=1.25em]{github.png}\hspace{.2em}
\href{https://github.com/JackMcKechnie/SARA}{\textsc{JackMcKechnie/SARA}}
\hspace{0.2em}
\includegraphics[width=1.25em,height=1.25em]{hf-logo.png}\hspace{.1em}
\href{https://huggingface.co/collections/JackMcKechnie/sara-indices}{\textsc{JackMcKechnie/sara-indices}}
\par}

\end{abstract}



\maketitle

\section{Introduction}
\looseness -1
Typical ad-hoc search tasks aim to provide a user with a subset of documents from a large corpus, ranked by their relevance to a query. As such, the task assumes that all of the information in the corpus can be seen by the user without causing harm. In scenarios such as web search, the documents in the corpus were written with the knowledge that they could be read by anyone. Therefore, \emph{sensitive} information such as personal information, medical diagnoses, or citizenship status is generally not included in web documents. However, many documents in archival collections, such as email collections \jm{or internal company documents}, were written with the expectation of a restricted or intended audience~\cite{email_audience}. Collections of such documents contain potentially sensitive information, intertwined with \jm{useful} non-sensitive information. \jm{Traditionally, to make such collections searchable, they must be} manually reviewed by expert-level sensitivity reviewers, \jm{however,} this is a prohibitively time-consuming and expensive process~\cite{graham_speed_up_1}.

\looseness -1
Such archival collections of documents can serve useful purposes if they are available to the public. For example, email collections can serve as a record of people's careers~\cite{gollins_review,gollins_review_2} or can document the operations of the company at a certain moment in time. \jm{Governments have identified that collections that potentially contain sensitive information, and therefore cannot currently be publicly accessed, contain information that could benefit stakeholders from a variety of sectors\jm{~\cite{gov_ai_sector}}. Access to such collections would bring economic and social benefits~\cite{gov_ai_sector,jaillant_call}.} Further, scholars in the Digital Humanities have called upon the Information Retrieval community~\cite{jaillant_call,jaillant_useful_1,jaillant_useful_2,jaillant_useful_3} to develop solutions that allow collections that potentially contain sensitive information to be made searchable. Many digital collections of documents have been donated to archives, but cannot be used by Digital Humanities researchers as they contain sensitive information amongst useful non-sensitive information. Previous work on making such collections available to the public has concentrated on \jm{the classification of sensitive information}. Models have been developed to automatically classify documents as sensitive or not~\cite{graham_towards,graham_word_embeddings,graham_kernerl} and have been deployed to speed up reviewer workflows~\cite{graham_speed_up_1,graham_speed_up_2}. However, there has been increasing interest in the literature on the development of \emph{sensitivity-aware search} (SAS) engines~\cite{sayed_jointly,oard_evaluating}, i.e., retrieval models that are able to search a collection \jm{that potentially contains} sensitive information and return documents that are relevant to the user query \jm{and are non-sensitive}~\cite{oard_evaluating}.

\looseness -1
The development of SAS models requires easily accessible test collections that comprise queries, a collection of documents (some that contain sensitive information and some that do not) with ground truth sensitivity assessments, and a set of query-document relevance assessments (qrels). Previous work has made some progress towards this, but each test collection comes with limitations. The TREC 2010 Legal Track~\cite{trec_2010_legal} provides a collection of legal documents that have been assessed by experts for relevance and legal privilege. Legal privilege is a specific type of sensitivity from e-discovery~\cite{oard_ediscovery_1,oard_ediscovery_2}, that allows information to be withheld for several specific reasons~\cite{oard_ediscovery_1}. However, for the TREC 2010 Legal track, documents were judged separately for relevance and sensitivity. Consequently, the overlap of documents that have been judged for both is small. Another collection that has made strides towards an easily accessible test collection for SAS is the Avocado collection~\cite{sayed_avocado}, which comprises documents from a now-defunct technology company and has been judged both for relevance to queries and for sensitivity. This is a useful collection for the development of SAS models. However, the Avocado collection is costly to obtain ($\sim$ \$1,500 USD at the time of writing), restricting its use to only those who can afford the fee. Further, the Avocado Collection is under a strict license, which limits the experiments that can be performed on it. For example, the documents cannot be shared, precluding experiments that involve user studies or crowdsourcing tasks.

\looseness -1
The Enron email corpus\jm{~\cite{introducing_enron}} is a free and publicly available collection of real emails that were released as part of an investigation by the Federal Energy Regulatory Commission into the collapse of the Enron Corporation. The corpus has gone through various iterations of cleaning, deduplication, and removal of emails upon request. Hearst et al.~\cite{hearst_enron} developed a rich taxonomy of categories that describe the contents of emails, based on a subset of the corpus. We refer to this subset as the Hearst Enron corpus. Examples of categories include press releases, internal company policy, and different emotional tones. The \emph{purely personal} and \emph{personal but in a professional context} categories are representative of sensitive personal information. Therefore, the Hearst Enron corpus contains documents and ground truth sensitivity labels, but does not contain queries or relevance judgements. In this work, we present an extension to the Enron email corpus that makes the corpus suitable for the development of sensitivity-aware search models. This is a collection of \textbf{S}ensitivity\jm{-}\textbf{A}ware \textbf{R}elevance \textbf{A}ssessments, as such we refer to the collection as the \emph{SARA} collection. We deploy a topic modelling approach to create 150 query formulations for 50 information needs. Further, we perform a crowdsourcing task to obtain 11,471 human-labelled relevance assessments for the Hearst Enron corpus. We also leverage automatic judgment techniques from the literature~\cite{umbrela,jack_context} to develop relevance assessments and sensitivity labels for all $\sim$500k documents of the full Enron corpus. \jm{We present baseline performance on the collection: (1) relevance performance of strong retrieval pipelines; (2) sensitivity classification performance using statistical models and three LLMs; and (3) SAS performance using post-retrieval classifier filtering of 3 retrieval models and 2 classifiers.} The \emph{SARA} collection is easily accessible via the ir\_datasets~\cite{ir_datasets} package. Additionally, to facilitate quick experimentation, we release sparse and dense indices for the collections as PyTerrier Artifacts~\cite{pyterrier_artifact}.

\looseness -1
The rest of this work is structured as follows: Section~\ref{sec:related_work} outlines relevant literature on sensitivity-aware search and similar tasks; Section~\ref{sec:existing_enron_collection} gives details on the existing Enron collection; Section~\ref{sec:development} describes how the collection is developed; Sections~\ref{sec:evaluation} and~\ref{sec:baselines} describe evaluation metrics for use with the collection and baseline performances, respectively; Section~\ref{sec:availability} gives details on how the collection can be easily accessed and experimented with \jm{and} briefly addresses possible ethical considerations with using the Enron email documents and Section~\ref{sec:conclusions} provides closing remarks.

\section{Related Work} \label{sec:related_work}

In this section, we discuss related work on sensitivity classification and existing sensitivity-aware search approaches.

\looseness -1
Gollins et al.~\cite{gollins_review} outlined the challenges that are associated with accessing collections of born-digital documents that potentially contain sensitive information. They note that the use of automatic sensitivity classifiers may be able to alleviate such issues. Subsequently, a line of work was carried out to develop sensitivity classifiers. For example, McDonald et al.~\cite{graham_towards,graham_word_embeddings} augmented SVM-based sensitivity classifiers with word embeddings and sensitivity-specific features to classify UK Freedom of Information Act~\cite{kamps_foia} (FOIA) sensitivities. Branting et al.~\cite{branting_sensitivity} developed a BERT-based model for US FOIA sensitivities, and Rollings et al.~\cite{rollings_foia} investigate the use of LLMs, also for US FOIA sensitivity classification.  However, whilst prior works were carried out on collections that contain ground truth sensitivity labels, such collections do not contain queries and relevance assessments, making them unsuitable for developing SAS approaches.

\jm{There has been recent interest in developing sensitivity-aware search approaches as a novel way to allow collections potentially containing sensitive information to be accessed~\cite{ecir_dc}.} Oard et al.~\cite{oard_evaluating} proposed a metric to evaluate SAS approaches, named Normalised Cost-Sensitive Discounted Cumulative Gain (nCSDCG). This metric rewards the system for returning relevant documents high in the ranking, but penalises the system for returning sensitive documents. Building upon nCSDCG, Sayed and Oard~\cite{sayed_jointly} proposed a learning-to-rank SAS approach that augments ranking features with sensitivity features. Additionally, McKechnie et al.~\cite{sans} proposed to train cross-encoder models for SAS by developing sensitivity-aware negative sampling methods. However, Sayed and Oard~\cite{sayed_jointly} and McKechnie et al.~\cite{sans} evaluated their approaches on the OHSUMED collection with a subset of PubMed Medical Subject Headings~\cite{mesh_labels} (categories of diseases) as a proxy for sensitive categories. This highlighted the need for a freely-available collection comprising queries, relevance assessments and ground truth sensitivity labels.

\looseness -1
To address the lack of appropriate collections for the development of SAS approaches, Sayed et al.~\cite{sayed_avocado} proposed an extension of the Avocado collection, comprising queries, relevance assessments, and sensitivity labels. The documents in the Avocado collection are emails from a now-defunct technology company, and the sensitivity labels are assigned based on two fictional personas~\cite{value_sensitive_design} that were developed to consider different risk tolerances. The Avocado collection provides a valuable resource for the development of sensitivity-aware search engines. However, the collection is expensive to obtain, prohibiting some researchers from using it, and is held under a strict license that precludes user studies, crowdsourcing, etc. 

\looseness -1
Finally, it is worth noting that, although Sayed et al.~\cite{sayed_jointly} introduced a collection that is suitable for use in developing sensitivity-aware search approaches, our \emph{SARA} extension of the Enron collection provides a valuable new resource to the field in two main ways. (1) The \emph{SARA} collection is made available to the community free of charge and without any prohibitive licensing arrangements. This makes our test collection suitable for exploratory research to evaluate ideas without substantial financial outlay, or to perform user studies or crowdsourcing experiments. Further, as our collection is available under an unrestrictive licence, we are able to release a number of artifacts that facilitate quick and easy experimentation. (2) Approaches that are developed on a narrowly defined set of sensitivities are unlikely to generalise. The Hearst Enron collection contains a range of classification categories that are representative of sensitive information\jm{, such as the \emph{purely personal} and \emph{personal but in a professional context} categories}. This provides opportunities to test \jm{the generalisability of SAS approaches}. Moreover, the \emph{SARA} collection can be used alongside the Avocado collection from Sayed et al.~\cite{sayed_avocado} to further test approaches developed on different collections.


\section{\jm{The} Enron Email Collection}~\label{sec:existing_enron_collection}
\looseness -1
The Enron Corporation was an American energy company that went bankrupt after it was revealed that the company had committed widespread accounting fraud~\cite{enron_downfall}. As part \jm{of} an investigation into the company, a corpus of $\sim$1.5 million emails sent and received by Enron employees was released. The collection was subsequently acquired by researchers at the Massachusetts Institute of Technology, and a curated version of $\sim$500k emails was made publicly available. \jm{After several iterations removing documents upon request from employees and further cleaning,} the current version of the Enron corpus is the 2015 version \jm{and is} released by Carnegie Mellon University (CMU)\jm{~\cite{introducing_enron}}.

\looseness -1
To be useful for developing sensitivity-aware search approaches, the corpus of documents must \jm{satisfy} the \jm{three} criteria that we outline here. (1) The corpus must discuss a range of topics, at least some of which must include sensitive information in the documents that are relevant to the topic. The Hearst~\cite{hearst_enron} subset of the Enron corpus contains 1702 documents that were annotated as relevant or not relevant to 53 different \jm{classification} categories. Some of these categories can reasonably be considered as representing true sensitive information, such as the \emph{Purely Personal} and \emph{Personal but in a Business Context} categories. \crc{We select these two categories to represent personal information sensitivities. However the collection also contains other categories, such as \emph{Humour} or \emph{Anxiety}, that could reasonably be expected to be potential signals of personal information.} (2) The collection must contain a diverse enough set of topics that a range of information needs can be developed. The nature of the Enron emails motivates their use as the emails were created in a real-world context, with employees of an actual company interacting naturally. Further, previous literature on topic modelling~\cite{enron_topic_modelling_1,enron_topic_modelling_2} has identified a rich and varied set of topics in the email collection. (3) Finally, the collection must be easily available and free of charge, to enable practitioners to develop sensitivity-aware search approaches with as low a barrier to entry as possible. The Enron collection is freely available and is not held under a restrictive licence; it is thus an appropriate collection to be extended to a SAS test collection.

\section{Development of the \textsc{SARA} Collection} \label{sec:development}

This section details how the \jm{\emph{SARA}} collection is built: pre-processing steps, crowdworking tasks, and automatic labelling approaches.

\looseness -1
\begin{sloppy}
\noindent\textbf{Document Preparation}
We use the CMU version\jm{~\cite{introducing_enron}} of the Enron collection comprising $\sim$500k documents. We remove duplicates via MD5 hashing~\cite{md5}. We take the body of the email as our document. Many emails carry essentially no information and are unlikely to be relevant to any information need. For example, some emails are lists of prices or \jm{are} one-word responses. As such, we deploy the QualT5~\cite{qualt5} model over the collection. QualT5 is typically used for index pruning to remove documents that are unlikely to be relevant to any topic. We use QualT5 to remove the 0.1\% lowest quality documents. \crc{We select this threshold via qualitative inspection of the documents.} Additionally, we hold out a set of 100,000 documents to be used as a training set for classifiers\jm{, to} ensure that there is no train-test leakage. The final \emph{SARA} \jm{collection} contains 129,821 documents.
\end{sloppy}

\looseness -1
\noindent\textbf{Information Needs}
To create our sensitivity-aware relevance assessments, we first identify themes that are present in the collection. We leverage these topics to develop descriptions of \jm{realistic} information needs that users might have, which can be satisfied by documents in the collection. We deploy a topic modelling approach to identify general themes that are covered in documents in the collection. Topics are selected to be broad enough that one could reasonably expect there to be relevant documents in the collection. We deploy topic modelling as it is an efficient way of identifying topics present in the collection via statistical modelling of the documents, rather than a blind search to manually identify topics.

\begin{figure}
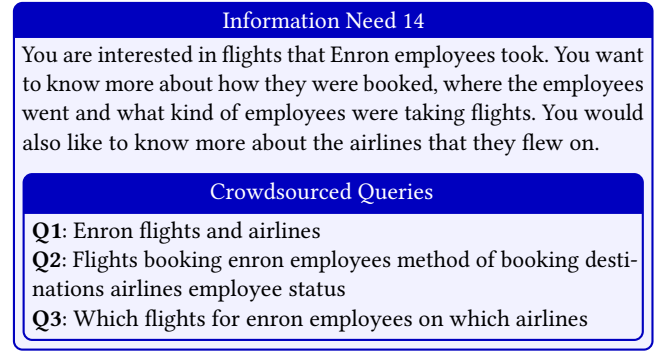

    \centering
    \observation{\phantom{111111111111111111}Information Need 14}{You are interested in flights that Enron employees took. You want to know more about how they were booked, where the employees went and what kind of employees were taking flights. You would also like to know more about the airlines that they flew on.

    \observation{\phantom{1111111111111111}Crowdsourced Queries}{\textbf{Q1}: Enron flights and airlines \\
\textbf{Q2}: Flights booking enron employees method of booking destinations airlines employee status \\
\textbf{Q3}: Which flights for enron employees on which airlines}}
    \caption{Example information need and queries.}
    \label{fig:info_need} 
\end{figure}

\looseness -1
We deploy Latent Dirichlet Allocation~\cite{lda} (LDA) topic modelling to identify 50 topics present in the collection. Subsequently, we perform manual sanity checks to ensure that there are relevant and sensitive documents in the collection for each topic. We use the top 10 terms from each topic (as identified by the LDA model) as queries, and use BM25~\cite{bm25} to search the collection. To be confident that there are relevant documents that are easily retrievable in the collection, we manually review the top $\sim$20 retrieved documents. Satisfied that there are relevant and sensitive documents present in the collection for each topic, we manually create short passages of text that serve as descriptions of the information needs that are to be satisfied by searching the collection. All of the information needs are of the same structure and are of similar size to the example \jm{presented} in Figure~\ref{fig:info_need}. The mean number of words in an information need passage is 41.48, with $\sigma^{2}$ = 27.37 words.

\looseness -1
\noindent\textbf{Query Formulations}
Typically, when a user has an information need, they formulate it as a textual query and issue it to a retrieval model. The retrieval model then aims to return a ranked list of documents to satisfy the information need of the user. Consequently, we require query formulations that represent our information needs. We collect query formulations for each topic from crowdworkers recruited on the Prolific crowdwork platform. Each crowdworker is shown 10 information needs and is asked to provide a query formulation that they would issue to a search engine to receive relevant documents to satisfy the information need. We recruit 10 crowdworkers for each batch of 10 information needs. We use attention checks to ensure that crowdworkers are engaged. Attention check questions are aligned with the guidelines of the Prolific platform. Crowdworkers are presented with a short piece of text that contains the correct (and randomised) answer to the question "What is your favourite colour?" and must select the correct option in a multiple-choice question. We discard those who do not answer the attention check correctly and reject their work on the Prolific platform. Crowdworkers are paid at a rate of £7/hour GBP\jm{, higher than the recommended compensation at the time of the experiment}. The output of this task is 10 different crowdsourced query formulations for each of the 50 information needs. \jm{Figure~\ref{fig:info_need} presents an illustrative example of these information needs and crowdsourced queries.}

\begin{table}[tb]
\caption{Rank correlation of 56 retrieval systems under nDCG@10 between the human and LLM judgements.}
\begin{tabular}{@{}llll@{}} 
\toprule
                    & Spearman $\rho$       & Kendall $\tau$        & $\tau_{AP}$           \\ \midrule
\textsc{SARA-human <> SARA-llm} & \multicolumn{1}{c}{0.8235} & \multicolumn{1}{c}{0.6154} & \multicolumn{1}{c}{0.5568} \\ \bottomrule
\end{tabular}
\label{tab:corr}
\end{table}

\noindent\textbf{Relevance Assessments}
To build an IR test collection, relevance labels for pairs of information needs and documents are required. We gather relevance assessments in two phases for the \emph{SARA} collection. First, we perform a crowdworking task to gather gold-standard human labels on the Hearst Enron collection\jm{~\cite{hearst_enron}}. Second, we augment the collection using approaches from the literature to develop LLM-generated relevance assessments. We perform analysis to compare human and LLM relevance assessments, shown in Section~\ref{sec:baselines}, \jm{finding high correlation of system rankings between sets of assessments}.

\noindent\textbf{Human Relevance Assessments} The Hearst Enron collection contains 1702 documents, and we have collected 50 information needs. To gather complete relevance assessments (85,100 judgements) would be financially infeasible~\cite{pooling}, therefore, we perform pooling to gather a subset of all \jm{query-document} pairs to be judged. We develop top-20 pools with 18 different systems; we use each combination of 3 \jm{queries selected from our crowdsourced query formulations}, 3 retrieval models (DPH~\cite{dph}, BM25~\cite{bm25}, PL2~\cite{pl2}), and using Bo1 query expansion~\cite{pl2} or not. The \jm{3} queries that we use are selected to provide a diverse set of documents, and after being read by the authors, they are deemed to be well formulated.

\looseness -1
We \jm{develop} a crowdsourcing task to collect relevance assessments for the query-document \jm{pairs in our pools}. Crowdworkers are shown an information need and an email, and are asked to \jm{label} the email as one of \emph{Highly Relevant}, \emph{Partially Relevant}, or \emph{Not Relevant} to the information need. If the crowdworker judges a document to be either \emph{Highly Relevant} or \emph{Partially Relevant}, then they are asked to select the passage of text that led them to their decision. We use this as an attention and quality check. In addition, we apply the same style of attention check as was used \jm{when collecting crowdsourced query formulations}. We pay crowdworkers at a rate of £7/hour. Each information need\jm{/}document pair is assessed by three crowdworkers, and we use a majority vote to decide the ground truth label. As we collect three judgments, and there are three possible labels, it is possible for there to be a tie. In practice, a tie only occurs for 134 information need\jm{/}document pairs, and we break ties by having one of the authors make an additional judgement. All crowdsourcing tasks were given ethical approval by the University of Glasgow College of Science and Engineering Ethics Committee \crc{(Approval No. 300220093)}.

\noindent\textbf{LLM Relevance Assessments} Collecting \jm{human judged} relevance assessments is an expensive and time-consuming process. To this end, we extend our crowdsourced human relevance assessments with LLM-generated relevance assessments. We \jm{develop} relevance assessments for our 150 crowdsourced query formulations to the full CMU Enron collection. As such, the qrels of the \emph{SARA} dataset include both human and LLM judgements. \jm{In spite of using an LLM to assess relevance, it is computationally and financially infeasible to judge $\sim$129k documents for 150 queries. As such,} we develop pools with lexical models (BM25, DPH, PL2, TFIDF w/ and w/o Bo1, KL, and RM3 query expansion), learned sparse models (SPLADE++, SPLADE-V2, SPLADE-V3), dense retrievers (TasB, ANCE, TCTColBERT, RetroMAE, E5), and cross-encoders (BM25 top 1000 reranked by monoT5, monoELECTRA, monoBERT, MiniLM), \crc{to} ensure varied pools. We perform top 1000 pooling. To collect relevance assessments, we deploy the \textsc{same query random document} approach proposed by McKechnie et al.~\cite{jack_context}. \jm{That is to say that we use a prompt based on the UMBRELA prompt~\cite{umbrela}, as this has been shown as effective in prior literature. Further, as per McKechnie et al.~\cite{jack_context}, when judging a query/document pair, we supply the model with few-shot examples that have been judged by a human for the same query. We use assessments from our crowdsourced relevance assessments as our human assessments.} We use the Llama 7B model~\cite{llama} as our LLM, \jm{as it has been shown to be effective and is efficient enough to practically be deployed on our large pools.}

\renewcommand{\arraystretch}{0.9}
\begin{table}[tb]
\caption{Baseline relevance performance on the \emph{SARA} dataset. We rerank the top 1000 documents from BM25. Results in \textbf{bold} indicate the best performance in the given metric, and \ul{underline} indicates second best.}
\resizebox{\linewidth}{!}{%
\begin{tabular}{@{}llcccc@{}}
\toprule
\phantom{1} & Approach                                                        & nDCG@10 &  RR & P@10  & R@100 \\ \midrule
\multirow{4}{*}{\rotatebox[origin=c]{90}{Lexical}} 
& BM25                                    & 0.5254  & 0.7044 & 0.5160 & 0.0511 \\
& TFIDF                                   & 0.5702  & 0.7291 & 0.5600 & 0.0544 \\
& PL2                                     & 0.5167  & 0.6923 & 0.5093 & 0.0503 \\
& DLH                                     & 0.5527  & 0.7211 & 0.5420 & 0.0526 \\ \midrule

\multirow{5}{*}{\rotatebox[origin=c]{90}{Dense}} 
& TasB                                & 0.5347  & 0.6759 & 0.5347 & 0.0581 \\
& ANCE                                & 0.5481  & 0.7201 & 0.5307 & 0.0484 \\
& TCTColBERT                          & 0.6161  & 0.7953 & 	0.5933 & 0.0570 \\
& RetroMAE                            & 0.5774  & 0.7784 & 0.5567 & 0.0503 \\
& E5                                  & 0.6300  & 0.8038 & 0.6147 & 0.0638 \\ \midrule

\multirow{3}{*}{\rotatebox[origin=c]{90}{Sparse}} 
& SPLADE++  &  0.6567 & 0.8073 & 0.6433 & 0.0614 \\
& SPLADE-V2 &  0.6390 & 0.7833 & 0.6273 & 0.0604 \\
& SPLADE-V3 &  0.6661 & 0.7950 & 0.6560 & 0.0634 \\ \midrule

\multirow{4}{*}{\rotatebox[origin=c]{90}{Cross-} \rotatebox[origin=c]{90}{Encoder}} 
& monoT5                             & 0.7002  & 0.8251 & 0.6887 & \ul{0.0721} \\
& monoELECTRA                        & 0.6872  & 0.8313 & 0.6733 & \textbf{0.0723} \\
& monoBERT                           & 0.6785  & 0.7974 & 0.6727 & 0.0672 \\
& MiniLM                             & 0.6666  & 0.8165 & 0.6513 & 0.0664 \\ \midrule

\multirow{3}{*}{\rotatebox[origin=c]{90}{LLM-} \rotatebox[origin=c]{90}{Rerank}} 
& RankVicuna-7B     & \ul{0.7161} & \ul{0.8642} & \ul{0.6933} & 0.0511 \\
& RankZephyr-7B     & \textbf{0.7429} & \textbf{0.9053} & \textbf{0.7087} & 0.0511 \\
& RankLlama-7B      & 0.6792 & 0.8203 & 0.6633 & 0.0660 \\ \bottomrule
\end{tabular}%
}
\label{tab:relevance}
\end{table}
\renewcommand{\arraystretch}{1}

\looseness -1
We follow best practices for using LLMs in IR evaluation, as set out by Dietz et al.~\cite{dietz_guidelines}. As such, we avoid the four undesirable evaluation tropes \jm{defined in~\cite{dietz_guidelines}}, which we explain here. \textbf{Eval. Trope \#1: Circularity} concerns leaking evaluation knowledge into the IR system. The retrieval systems that we deploy as baselines have no knowledge of the evaluation criteria. \textbf{Eval. Trope \#2: LLM Evaluator as a Ranker} concerns the use of the same model for ranking and evaluation. Our relevance assessments include human and LLM assessments; the same model is not used for ranking and evaluation. Further, we test RankLlama~\cite{rankllama} (a ranker based on the same model as our evaluator) and it is not our highest-performing baseline. \crc{It should be noted that deploying the labelling approach proposed by McKechnie et al.~\cite{jack_context} as a ranker would be problematic, as the roles of ranker and evaluator would be conflated. That is to say, systems that are tested on the \emph{SARA} dataset should not leverage knowledge of how relevance assessments are collected to improve performance.} \textbf{Eval. Trope \#3: LLM Narcissism} posits that LLMs prefer text that was generated by LLMs over text written by humans. The Enron documents are emails written by humans, and the company ceased trading in 2007; as such, LLM Narcissism is not \jm{a} concern. Finally, \textbf{Eval. Trope 4: Loss of Variety of Opinion} pertains to the loss of different points of view when a single model is used for relevance assessment, rather than many different human assessors. The \emph{SARA} collection includes relevance assessments from multiple crowdworkers and from an LLM, hence there is a variety of opinion.


\noindent\textbf{Sensitivity Labels} We extend the Enron collection with sensitivity labels for all \jm{129,821} documents, to complement the manually sensitivity reviewed documents in Hearst subset. To do so, we train a T5-based model on the Hearst \jm{Enron Subset to label document as either \emph{purely personal/personal but in a business context} (sensitive) or \emph{not personal} (non-sensitive)}, \jm{and deploy the model on} the unlabelled documents. We train our model using 5-fold cross-validation, and it achieves Precision: 0.4667, Recall: 0.5308, Accuracy: 0.8666, F1: 0.4967 Balanced Accuracy: 0.7225 averaged across the folds. Further details on the \crc{distribution of the} sensitivity labels that we release as part of the \emph{SARA} collection are given in Section~\ref{sec:baselines}.

\section{Evaluation Metrics} \label{sec:evaluation}
\looseness -1
There are three aspects of sensitivity-aware search that can be evaluated; (1) the relevance of the top-$k$ documents that the model retrieves, independent \jm{from whether} they are sensitive or not; (2) the amount of sensitive information that is contained in the top-$k$ documents, regardless of relevance to the query; (3) the tradeoff between \jm{the} relevance and sensitivity of the top-$k$ \jm{documents}. We make recommendations on metrics to use to evaluate these three aspects.

\looseness -1
To evaluate the relevance of results, we recommend nDCG~\cite{ndcg} as both the relevance of results and their position are considered, and nDCG has been used in SAS literature~\cite{oard_evaluating,sayed_jointly,sans}. We recommend sens\_docs@\jm{$k$} to measure the sensitivity of the top-$k$ results. sens\_docs\jm{@k} measures the average number of documents that contain sensitive information, averaged across queries, and is defined as 
\[
\text{sens\_docs@}k(q)
= \sum_{i=0}^{\jm{k}} \mathbbm{1}\!\{d_{i} \in \mathcal{S}\}
\]
where $S$ is the set of documents in the collection that contain sensitive information. Finally, to measure the relevance versus sensitivity tradeoff, we recommend the use of the \emph{cost-sensitive} nDCG~\cite{oard_evaluating} (nCSDCG) metric, which rewards the system for retrieving relevant documents and penalises the system for returning sensitive documents. nCSDCG is the min-max normalisation of 
\[
\sum_{i=0}^{k} \frac{g_{i}}{d_{i}} - c_i
\] 
where $g_{i}$ is a gain value and $d_{i}$ is a discount value, both set according to the nDCG metric. $c_{i}$ is a cost penalty for retrieving sensitive documents. Hence, $c_{i}$ should be 0 when the document is non-sensitive and $>0$ when the document is sensitive. We recommend nCSDCG@\jm{$k$} is used to evaluate the relevance versus sensitivity tradeoff, and that the cost penalty $c_{i}$ is set to the maximum DCG value, following previous SAS literature\jm{~\cite{oard_evaluating,sayed_avocado,sans}}.

\begin{table}[tb]
\caption{Classification baselines on the \emph{SARA} dataset.}
\resizebox{\linewidth}{!}{
\begin{tabular}{@{}lccccc@{}}
\toprule
Approach          & \multicolumn{1}{c}{Prec.} & \multicolumn{1}{c}{Recall} & \multicolumn{1}{c}{Acc.} & \multicolumn{1}{c}{F1} & \multicolumn{1}{c}{BAC} \\ \midrule
LR                & 0.4454                            & \ul{0.7147}                          & 0.8433                            & 0.5488                      & 0.7889                                      \\
SVM               & \ul{0.4469}                             & 0.7141                          & \ul{0.8440}                            &  \ul{0.5497}                     & \ul{0.7891}                                      \\
T5                & \textbf{0.6038}                            & \textbf{0.7625}                          & \textbf{0.9016}                            & \textbf{0.6740}                      & \textbf{0.8428}                                      \\
Zero-Shot Llama 8B   & 0.1712                             & 0.1387                          & 0.7956                            & 0.1533                      & 0.5177                                      \\
Few-Shot Llama 8B    & 0.1595                             & 0.4313                          & 0.6153                            & 0.2328                       & 0.5377                                      \\
Zero-Shot Aya-Expanse 32B       & 0.2475                             & 0.2095                          & 0.8096                            & 0.2269                      & 0.5558                                      \\
Few-Shot Aya-Expanse 32B        & 0.2403                             & 0.2594                          & 0.7919                            & 0.2495                      & 0.5666                                      \\ 
Zero-Shot Qwen 72B & 0.2092                             & 0.3265                          & 0.7493                             & 0.2550                      & 0.5699                                      \\
Few-Shot Qwen 72B  & 0.2092                             & 0.3265                          & 0.7493                            & 0.2550                      & 0.5699                                      \\ \bottomrule
\end{tabular}%
} 
\label{tab:classification}
\end{table}
\section{Collection Analysis and Baselines} \label{sec:baselines}

In this section, we \jm{present a} detailed analysis of the \emph{SARA} collection and provide baseline performances for sensitivity classification, relevance-only retrieval, and sensitivity-aware search.

\noindent\textbf{Collection Analysis} The \emph{SARA} collection comprises 129,821 documents, of which 112,510 are non-sensitive, and 17311 are sensitive ($\sim$15\%). Other collections \jm{with potentially sensitive information} contain similar levels of sensitive information~\cite{graham_towards,sayed_avocado}. 100,000 documents are used as \jm{a} held\jm{-}out training set, \crc{13326 of which are sensitive and 86674 of which are non-sensitive ($\sim$13\% sensitive)}. 95,960 unique documents have been judged for 150 query formulations (50 topics), resulting in  800,157 qrels (11,471 human judged). Each query has on average 5334 ($\sigma$=1005) judged documents, of which 1034 ($\sigma$=780) are relevant. Each query has, on average, 979 ($\sigma$=742) relevant non-sensitive documents and 57 ($\sigma$=54) relevant sensitive documents.

\looseness -1
In a typical IR evaluation setup, we aim to determine the \emph{ordering} of systems\jm{,} to decide which are better than others~\cite{voorhees_variations}. As such, we demonstrate the validity of our LLM judged relevance assessments by comparing the ranking of a set of systems under our human relevance assessments with the ranking of the same set of systems \jm{under} our LLM relevance assessments. To this end, we evaluate the following systems using the nDCG@10~\cite{ndcg} metric with human and LLM relevance assessments: BM25~\cite{bm25}, DPH~\cite{dph}, PL2~\cite{pl2},  TFIDF~\cite{tfidf} (with and without Bo1~\cite{pl2}, KL~\cite{pl2}, and RM3~\cite{rm3} query expansion) ANCE~\cite{ance}, RetroMAE~\cite{retromae}, TasB~\cite{tasb}, TCTColBERT~\cite{tctcolbert}, E5~\cite{e5}, SPLADE++~\cite{splade_pp}, SPLADE-v2~\cite{splade_v2}, SPLADE-v3~\cite{splade_v3}, and the top 1000 documents from BM25, DPH, PL2, and TFIDF reranked by monoT5~\cite{monot5}, miniLM~\cite{minilm}, ELECTRA~\cite{electra}, monoBERT~\cite{monobert}. We then calculate Spearman $\rho$~\cite{spearman_rho}, Kendall $\tau$~\cite{kendall_tau}. and $\tau_{AP}$~\cite{tau_ap} rank correlation metrics. We use $\tau_{AP}$ as it is a top-heavy rank correlation measure. The results of this experiment are shown in Table~\ref{tab:corr}. We observe that our LLM relevance assessments are effective, as their ranking of retrieval systems is highly correlated with the ranking under human assessments with $> 0.8$ Spearman $\rho$ and $>0.6$ Kendall $\tau$, agreement levels in line with previous studies~\cite{faggioli_llm,umbrela,chuan_judges}. As such, we have confidence that our LLM relevance assessments are valid.

\begin{table}[tb]
\caption{Baseline SAS performance on the \emph{SARA} dataset.}
\resizebox{\linewidth}{!}{%
\begin{tabular}{@{}lccc@{}}
\toprule
Approach                                               & \multicolumn{1}{l}{nDCG@10} & \multicolumn{1}{l}{sens\_docs@10} & \multicolumn{1}{l}{nCSDCG@10} \\ \midrule
BM25 & 0.5254 &	0.7467	& 0.9046 \\
BM25 $\gg$ Llama ZS Filter                    & 0.5201                           & 0.7200                                 & 0.9069                             \\ 
BM25 $\gg$ T5 Filter                     & 0.5281                           & 0.2933                                 & 0.9477                             \\ \midrule

BM25 $\gg$ monoT5 & 0.7002 &	0.4533 &	0.9415 \\
BM25 $\gg$ monoT5 $\gg$ Llama ZS Filter       & 0.6962                           & 0.3800                                 & 0.9482                             \\ 
BM25 $\gg$ monoT5 $\gg$ T5 Filter        & 0.7030	                           & 0.1667	                                 & 0.9688                             \\ \midrule

BM25 $\gg$ RankZephyr-7B & 0.7429 &	0.8133 &	0.9096 \\
BM25 $\gg$ RankZephyr-7B $\gg$ Llama ZS Filter & 0.7296                           & 0.7600                                 & 0.9139                             \\ 
BM25 $\gg$ RankZephyr-7B  $\gg$ T5 Filter  & 0.7295                           & 0.2800	                                 & 0.9594                             \\ \bottomrule
\end{tabular}%
}
\label{tab:sas}
\end{table}

\looseness -1
\crc{\textbf{Qualitative Examples} We now provide examples of sensitive and non-sensitive documents from the \emph{SARA} collection. We describe the properties that make such documents (non-)sensitive, and why the sensitivity of some documents is more difficult to identify than others. Figure~\ref{fig:example_docs} shows such example documents. Example E1 (the top document) is an example of a non-sensitive document since it discusses matters that are purely related to the operation of the Enron corporation. Mentioned in the document are the banking matters of ECT and ENA, which are subsidiaries of the Enron Corporation. No other topic is discussed, and as such, the document is non-sensitive. No personal information is discussed. Turning to Example E2, we note mentions of spouses, personal calls, a medical diagnosis, and family connections. Nowhere in Example E2 does it discuss business matters that are related to Enron. Consequently, we can see that Example E2 is a sensitive document since it discusses purely personal information. Moreover, it is likely that Example E2 is not a challenging document to identify as sensitive because: (1) there is a single, clearly personal topic discussed; and (2) entities such as spouses and children are mentioned, which are not typical in business emails. Finally, we turn to Example E3, which discusses the work experience of a prospective employee of Enron, named Jason. Mick McConnell, President and Chief Executive Officer for Enron Global Markets, recommends Jason for a job and wishes him well. We can see that Example E3 is professional in some senses, i.e. the example discusses employment, but it is also personal as Jason and Mick clearly know each other, and Mick is helping Jason. As such, Example E3 is an example of a document that is personal, but in a professional context. We can observe that Example E3 is likely to be more difficult to identify as sensitive, as it both discusses employment and has some personal elements to it. Through these three example documents from the \emph{SARA} collection, we can see the different types of documents present in the collection, and what makes them likely to be difficult (or not) to identify as sensitive.}

\begin{figure}
    \centering
    \nonsensitivedoc{Example E1 - Non-Sensitive Document}{Sara, Are you certain that the executing broker - equities list is definitive? I reviewed the list of executing brokers that are currently used by ECTI and ENA, and at least half were not on this list. Is it possible that we are trading without agreements with all of these brokers? examples: Lehman Brothers, Hambrecht \& Quist, Thomas Weisel Partners, DLJ, SBC Warburg Dillon Read to name a few. Could you have someone check on this? Thanks!}
        
    \sensitivedoc{Example E2 - Sensitive Document}{Mike, ? You will probably receive a call from the top JDRF Legislative Volunteer, Leah Mullin. Her husband, Leo Mullin, is CEO of Delta Airlines and they have a diabetic son. They are also very friendly with the Brennemans. ? Leah would like to enlist the help of yourself and any Enron powers-that-be for support from Dick Cheney for stem cell funding. ? As you know, from our many mailouts, this is a hot topic! We are very afraid that Pres Bush and VP Cheney might ban stem cell research, thus negating all chances of our cure for diabetes. ? I hope that you can help Leah when she calls. I have sung your praises to her (and just about everyone from Washington to Timbucktu). Thanks, Mike. Kathy}

    \sensitivedoc{Example E3 - Sensitive Document}{Jason, I have forwarded your resume with my commentary to Philippe Bibi, CEO of Enron Networks and Robert Jones, VP of HR for ENW. You have an impressive resume and I hope you get a good response from Enron. Best of luck, Mike McConnell}
    \caption{Examples of sensitive documents from \emph{SARA}. Shown are a non-sensitive document (top), a purely personal document (middle) and a document that is personal but in a professional context (bottom).}
    \label{fig:example_docs}
\end{figure}

\noindent\textbf{Baseline Performance} We \crc{now present baseline performances on the \emph{SARA} collection, starting with} relevance-only baselines. Table~\ref{tab:relevance} shows the relevance-only performance of different families of retrieval models. We experiment with lexical, dense, and sparse retrievers, cross-encoders and LLM-based rerankers from the literature. Results follow the general trends that \jm{one} would expect from the literature. Dense retrievers outperform lexical retrievers (e.g. 0.5254 $\rightarrow$ 0.6161 nDCG@10, BM25 vs. TCTColBERT) and sparse retrievers outperform dense retrievers (e.g. 0.6300 $\rightarrow$ 0.6567 nDCG@10, E5 vs. SPLADE++). Reranking using cross-encoders and LLM-based models provides further improvements (e.g. 0.6661 $\rightarrow$ 0.7002, 0.7429 nDCG@10, SPLADE-V3 vs. BM25 $\gg$ monoT5 and BM25 $\gg$ RankZephyr-7B, respectively). In conclusion, we demonstrate that reasonable relevance performance that follows trends from the literature is achievable on the \emph{SARA} collection.

\looseness -1
SAS requires the retrieval model to be able to both rank documents for their relevance and to \crc{automatically} identify sensitive information. As such, it must be possible to effectively identify the sensitive information that is included in a SAS test collection\crc{, with automated methods}. Hence, we demonstrate baseline classification performance on the \emph{SARA} collection. We train logistic regression (LR), support vector machine (SVM) and T5-based classifiers on the \emph{SARA} training set. Further, we prompt three LLMs zero-shot and with 3 randomly selected few-shot examples. We use the \textsc{llama-3-8b-instruct}~\cite{llama}, \textsc{aya-expanse-32b}~\cite{aya}, and \textsc{qwen-2.5-72b-instruct}~\cite{qwen} models to capture a variety of different families and sizes of LLM. Results of this experiment are shown in Table~\ref{tab:classification}. We observe that the T5-based classifier outperforms the classical baselines of LR and SVM (e.g. 0.7889 $\rightarrow$ 0.5428 BAC, LR vs. T5). Additionally, we see that\jm{,} in general, few-shot approaches outperform zero-shot approaches and larger LLMs outperform smaller ones (e.g. 0.5177 $\rightarrow$ 0.5699 \jm{BAC} Zero-Shot Llama 8B vs. Zero-Shot Qwen 72B). As such, \jm{we} conclude that reasonable classifier baselines are able to effectively identify the sensitive information in the \emph{SARA} collection. \crc{We have hence established that the sensitive information that is contained in the \emph{SARA} can be automatically identified, a key requirement for a sound SAS test collection.} Indeed, the collection can be used as a stand-alone classification dataset.

\begin{figure}
\begin{lstlisting}[language=Python]
import pyterrier as pt
from ir_measures import *
# Load the dataset and index
dataset = pt.get_dataset('irds:sara')
index = pt.Artifact.from_hf("JackMcKechnie/sara.terrier")
# Define retrievers
tfidf = index.tf_idf()
bm25 = index.bm25()
# Define sens_docs measure
docno2sens = dict(dataset.get_corpus_iter()[["docno", "sensitivity"]].values)
sens_docs = define_byquery(
    lambda qrels, run: run["docno"].map(docno2sens).sum(),
    name="sens_docs"
)
# Run PyTerrier experiment!
pt.Experiment(
    [tfidf, bm25],
    dataset.get_topics(),
    dataset.get_qrels(),
    eval_metrics=[nDCG@10, sens_docs@10]
)
\end{lstlisting}
\caption{An example of accessing the \emph{SARA} collection via ir\_datasets and performing a PyTerrier experiment.}
\label{fig:code}
\end{figure}

\looseness -1
Finally, we provide post-filtering SAS baseline performances. We use the best and the worst classifiers from our previous experiments as a post-retrieval filter~\cite{sayed_jointly} over three retrieval pipelines. This represents a simple and intuitive SAS baseline. We evaluate using nDCG@10, sens\_docs@10, and nCSDCG@10 and present experimental results in Table~\ref{tab:sas}. We observe that the Zero-Shot Llama (Llama ZS) filter is ineffective at removing sensitive documents from the retrieved results (e.g. 0.7467 $\rightarrow$ 0.7200 sens\_docs@10, BM25 vs. BM25 $\gg$ Llama ZS Filter), which is reflected in the relevance vs. sensitivity tradeoff (0.9046 $\gg$ 0.9069 nCSDCG@10, BM25 vs. BM25 $\gg$ Llama ZS Filter). However, when the more effective T5 classifier is applied, fewer sensitive documents are in the top 10 \jm{results}, and a better tradeoff is achieved. Further, improvements in relevance with similar levels of sensitivity are reflected in the relevance vs. sensitivity tradeoff (e.g. 0.9477 $\rightarrow$ 0.9594 nCSDCG@10, BM25 $\gg$ T5 Filter BM25 $\gg$ RankZephyr-7B $\gg$ T5 Filter). To this end, we have shown reasonable SAS baseline performance on the \emph{SARA} dataset.

\section{Availability and Ethical Considerations} \label{sec:availability}

\looseness -1
The \emph{SARA} dataset is held under an AttributionNonCommercial 4.0 International licence, which
allows for it to be adapted, transformed and built upon. We make documents, queries, relevance assessments, and sensitivity judgements available through the popular ir\_datasets~\cite{ir_datasets} package. We present an example that shows how access the \emph{SARA} dataset and perform an experiment in Figure~\ref{fig:code}. Furthermore, we release pre-built indices\footnote{\hflink{https://huggingface.co/collections/JackMcKechnie/sara-indices}{JackMcKechnie/sara-indices}} to allow the use of different weighting models and dense retrieval models. These indices can be easily used with PyTerrier Artifacts~\cite{pyterrier_artifact}. By releasing indices, we enable researchers to perform experiments using effective sparse and dense retrievers, without the need for expensive GPU computation. Additionally, we release the dataset as CSV files on GitHub.\footnote{\githublink{https://github.com/JackMcKechnie/SARA}{JackMcKechnie/SARA}}

\looseness 1

\looseness -1
\noindent\textbf{Ethical Considerations}
Previous work has had concerns about using the Enron Email Collection for the creation of a dataset for sensitivity-aware search~\cite{sayed_avocado}, which we address in this section. Sayed et al.~\cite{sayed_avocado} argued that the ultimate goal of sensitivity-aware search is to protect sensitive information. Therefore, in their opinion, using crowdworkers to annotate a corpus which contains sensitive information goes against the goal of protecting \jm{such} information. However, the Enron Email Collection has been publicly available for over a decade and has undergone numerous redaction efforts. Therefore, those ex-Enron employees who felt that they did not want their emails to be read by others have an opportunity over the last 10 years to request the removal of their emails. Additionally, the emails that are shown to the crowdworkers to create our relevance assessments are publicly available and widely used. Consequently, we argue that the use of the Enron Email Collection to build a sensitivity-aware search test collection is justified and provides a valuable resource
for the community\jm{, due to the difficulty of obtaining and distributing any other collection with sensitivities.}

\section{Conclusions} \label{sec:conclusions}

In this work, we have identified the need for test collections that are suitable for the development of sensitivity-aware search retrieval models. We \jm{have} presented our extension to the Enron email collection that includes queries, relevance assessments, and sensitivity judgements for the existing Enron documents. Topic modelling was carried out to identify topics present in the Enron email collection. Based on these topics, 50 information needs were developed. Crowdsourcing tasks were carried out to collect 150 query formulations and query-document relevance assessments. This was extended with relevant assessments and sensitivity judgments from a large language model. We showed baseline performance for relevance-only, sensitivity classification, and sensitivity-aware search on the collection. Our experiments illustrate the usefulness of our sensitivity-aware relevance assessments extension to the Enron email collection for evaluating sensitivity-aware search systems.

\balance
\bibliographystyle{ACM-Reference-Format}
\bibliography{bib.bib}

@inproceedings{gollins_review,
  title={On Using Information Retrieval for the Selection and Sensitivity Review of Digital Public Records},
  author={Gollins, Timothy and McDonald, Graham and Macdonald, Craig and Ounis, Iadh},
  booktitle={PIR @ SIGIR},
  year={2014}
}

@article{gollins_review_2,
  title={Our digital legacy: an archival perspective},
  author={Moss, Michael S and Gollins, Tim J},
  journal={JCAS},
  volume={4},
  number={2},
  year={2017}
}

@article{jaillant_useful_1,
  title={How can we make born-digital and digitised archives more accessible? Identifying obstacles and solutions},
  author={Jaillant, Lise},
  journal={Arch. Sci.},
  volume={22},
  number={3},
  year={2022}
}

@article{jaillant_useful_2,
  title={Applying AI to digital archives: trust, collaboration and shared professional ethics},
  author={Jaillant, Lise and Rees, Arran},
  journal={DSH},
  volume={38},
  number={2},
  year={2023},
}

@book{jaillant_useful_3,
  title={Archives, access and artificial intelligence: working with born-digital and digitized archival collections},
  author={Jaillant, Lise},
  year={2022},
  publisher={Bielefeld University Press}
}

@article{jaillant_call,
  title={Unlocking digital archives: cross-disciplinary perspectives on AI and born-digital data},
  author={Jaillant, Lise and Caputo, Annalina},
  journal={AI Soc},
  volume={37},
  number={3},
  year={2022}
}

@inproceedings{graham_towards,
  title={Towards a classifier for digital sensitivity review},
  author={McDonald, Graham and Macdonald, Craig and Ounis, Iadh and Gollins, Timothy},
  booktitle={Proc. of ECIR},
  year={2014}
}

@inproceedings{graham_word_embeddings,
  title={Enhancing sensitivity classification with semantic features using word embeddings},
  author={McDonald, Graham and Macdonald, Craig and Ounis, Iadh},
  booktitle={Proc. of ECIR},
  year={2017}
}

@inproceedings{graham_kernerl,
  title={A study of SVM kernel functions for sensitivity classification ensembles with POS sequences},
  author={McDonald, Graham and Garc{\'\i}a-Pedrajas, Nicol{\'a}s and Macdonald, Craig and Ounis, Iadh},
  booktitle={Proc. of SIGIR},
  year={2017}
}

@inproceedings{graham_speed_up_1,
  title={Towards maximising openness in digital sensitivity review using reviewing time predictions},
  author={McDonald, Graham and Macdonald, Craig and Ounis, Iadh},
  booktitle={Proc. of ECIR},
  year={2018},
}

@article{graham_speed_up_2,
  title={How the accuracy and confidence of sensitivity classification affects digital sensitivity review},
  author={McDonald, Graham and Macdonald, Craig and Ounis, Iadh},
  journal={TOIS},
  volume={39},
  number={1},
  year={2020}
}

@inproceedings{sayed_jointly,
  title={Jointly modeling relevance and sensitivity for search among sensitive content},
  author={Sayed, Mahmoud F and Oard, Douglas W},
  booktitle={Proc. of SIGIR},
  year={2019}
}

@inproceedings{oard_evaluating,
  title={Evaluating Search Among Secrets},
  author={Oard, Douglas W. and Shilton, Katie and Lin, Jimmy},
  booktitle={EVIA @ NTCIR},
  year={2016}
}

@inproceedings{trec_2010_legal,
  title={Overview of the TREC 2010 legal track},
  author={Cormack, Gordon V and Grossman, Maura R and Hedin, Bruce and Oard, Douglas W.},
  booktitle={Proc. of TREC},
  year={2010}
}

@article{oard_ediscovery_1,
  title={Jointly minimizing the expected costs of review for responsiveness and privilege in e-discovery},
  author={Oard, Douglas W. and Sebastiani, Fabrizio and Vinjumur, Jyothi K},
  journal={TOIS},
  volume={37},
  number={1},
  year={2018},
}

@article{oard_ediscovery_2,
  title={Finding the privileged few: Supporting privilege review for e-discovery},
  author={Vinjumur, Jyothi K and Oard, Douglas W.},
  journal={Proc. of ASIS\&T},
  volume={52},
  number={1},
  year={2015},
}

@inproceedings{sayed_avocado,
  title={A test collection for relevance and sensitivity},
  author={Sayed, Mahmoud F and Cox, William and Rivera, Jonah Lynn and Christian-Lamb, Caitlin and Iqbal, Modassir and Oard, Douglas W. and Shilton, Katie},
  booktitle={Proc. of SIGIR},
  year={2020}
}

@inproceedings{hearst_enron,
  title={Teaching applied natural language processing: Triumphs and tribulations},
  author={Hearst, Marti A},
  booktitle={ETMTNLP @ ACL},
  year={2005}
}

@article{umbrela,
  title={Umbrela: Umbrela is the (open-source reproduction of the) bing relevance assessor},
  author={Upadhyay, Shivani and Pradeep, Ronak and Thakur, Nandan and Craswell, Nick and Lin, Jimmy},
  journal={arXiv preprint arXiv:2406.06519},
  year={2024}
}

@inproceedings{jack_context,
  title={Context Example Selection for LLM Generated Relevance Assessments},
  author={McKechnie, Jack and McDonald, Graham and Macdonald, Craig},
  booktitle={Proc. of ECIR},
  year={2025}
}

@inproceedings{ir_datasets,
  title={Simplified data wrangling with ir\_datasets},
  author={MacAvaney, Sean and Yates, Andrew and Feldman, Sergey and Downey, Doug and Cohan, Arman and Goharian, Nazli},
  booktitle={Proc. of SIGIR},
  year={2021}
}

@inproceedings{pyterrier_artifact,
  title={Artifact Sharing for Information Retrieval Research},
  author={MacAvaney, Sean},
  booktitle={Proc. of SIGIR},
  year={2025}
}

@article{branting_sensitivity,
  title={Decision support for detecting sensitive text in government records: Anonymous submission},
  author={Branting, Karl and Brown, Bradford and Giannella, Chris and Guilder, James Van and Harrold, Jeff and Howell, Sarah and Baron, Jason R},
  journal={AI Law},
  volume={33},
  number={1},
  year={2025},
}

@inproceedings{rollings_foia,
  title={Using ChatGPT for the FOIA Exemption 5 Deliberative Process Privilege},
  author={Baron, Jason R and Rollings, Nathaniel W and Oard, Douglas W.},
  booktitle={LegalAIIA @ ICAIL},
  year={2023}
}

@article{ndcg,
  title={Cumulated gain-based evaluation of IR techniques},
  author={J{\"a}rvelin, Kalervo and Kek{\"a}l{\"a}inen, Jaana},
  journal={TOIS},
  volume={20},
  number={4},
  year={2002},
}

@inproceedings{sans,
  title={Bi-Objective Negative Sampling for Sensitivity-Aware Search},
  author={McKechnie, Jack and McDonald, Graham and Macdonald, Craig},
  booktitle={Proc. of SIGIR},
  year={2024}
}

@article{mesh_labels,
  title={Medical subject headings (MeSH)},
  author={Lipscomb, Carolyn E},
  journal={BMLA},
  volume={88},
  number={3},
  year={2000}
}

@article{value_sensitive_design,
  title={Search with discretion: value sensitive design of training data for information retrieval},
  author={Iqbal, Modassir and Shilton, Katie and Sayed, Mahmoud F and Oard, Douglas and Rivera, Jonah Lynn and Cox, William},
  journal={Proc. of PACMHCI},
  volume={5},
  year={2021},
}

@article{enron_downfall,
  title={The rise and fall of Enron},
  author={Thomas, C William},
  journal={J. Account},
  volume={193},
  number={4},
  year={2002},
}

@inproceedings{enron_topic_modelling_1,
  title={Topic modeling in the enron dataset},
  author={Celebi, Naciye and Shashidhar, Narasimha},
  booktitle={Proc. of BigData},
  year={2022},
}

@inproceedings{enron_topic_modelling_2,
  title={The author-recipient-topic model for topic and role discovery in social networks: Experiments with enron and academic email},
  author={McCallum, Andrew and Corrada-Emmanuel, Andr{\'e}s and Wang, Xuerui},
  year={2004},
  booktitle={NIPS’04 Workshop on Structured Data and Representations in Probabilistic Models for Categorization}
}

@misc{md5,
  title={RFC1321: The MD5 message-digest algorithm},
  author={Rivest, Ronald},
  year={1992},
}

@inproceedings{qualt5,
  title={Neural Passage Quality Estimation for Static Pruning},
  author={Chang, Xuejun and Mishra, Debabrata and Macdonald, Craig and MacAvaney, Sean},
  booktitle={Proc. of SIGIR},
  year={2024}
}

@article{bm25,
  title={A probabilistic model of information retrieval: development and comparative experiments: Part 2},
  author={Jones, Karen Sparck and Walker, Steve and Robertson, Stephen E.},
  volume={36},
  number={6},
  year={2000},
}

@article{pooling,
  title={Information retrieval test collections},
  author={Sparck Jones, Karen and Van Rijsbergen, Cornelis Joost},
  journal={J. Doc.},
  volume={32},
  number={1},
  year={1976}
}

@inproceedings{dph,
  title={{FUB}, {IASI-CNR} and University of Tor Vergata at {TREC} 2008 blog track},
  author={Amati, Giambattista and Amodeo, Giuseppe and Bianchi, Marco and Gaibisso, Carlo and Gambosi, Giorgio},
  booktitle={Proc. of TREC},
  year={2008}
}

@article{pl2,
  title={Probabilistic models of information retrieval based on measuring the divergence from randomness},
  author={Amati, Gianni and Van Rijsbergen, Cornelis Joost},
  journal={TOIS},
  volume={20},
  number={4},
  year={2002}
}

@article{llama,
  title={Llama: Open and efficient foundation language models},
  author={Touvron, Hugo and Lavril, Thibaut and Izacard, Gautier and Martinet, Xavier and Lachaux, Marie-Anne and Lacroix, Timoth{\'e}e and Rozi{\`e}re, Baptiste and Goyal, Naman and Hambro, Eric and Azhar, Faisal and others},
  journal={arXiv preprint arXiv:2302.13971},
  year={2023}
}

@inproceedings{dietz_guidelines,
  title={Principles and Guidelines for the Use of LLM Judges},
  author={Dietz, Laura and Zendel, Oleg and Bailey, Peter and Clarke, Charles LA and Cotterill, Ellese and Dalton, Jeff and Hasibi, Faegheh and Sanderson, Mark and Craswell, Nick},
  booktitle={Proc. of ICTIR},
  year={2025}
}

@inproceedings{rankllama,
  title={Fine-tuning llama for multi-stage text retrieval},
  author={Ma, Xueguang and Wang, Liang and Yang, Nan and Wei, Furu and Lin, Jimmy},
  booktitle={Proc. of SIGIR},
  year={2024}
}

@inproceedings{voorhees_variations,
  title={Variations in relevance judgments and the measurement of retrieval effectiveness},
  author={Voorhees, Ellen M},
  booktitle={Proc. of SIGIR},
  year={1998}
}

@article{tfidf,
  title={A statistical interpretation of term specificity and its application in retrieval},
  author={Sparck Jones, Karen},
  journal={J. Doc},
  volume={28},
  number={1},
  year={1972}
}

@inproceedings{rm3,
  title={UMass at {TREC} 2004: Novelty and HARD},
  author={Abdul-Jaleel, Nasreen and Allan, James and Croft, W Bruce and Diaz, Fernando and Larkey, Leah and Li, Xiaoyan and Smucker, Mark D and Wade, Courtney},
  booktitle={Proc. of TREC},
  year={2004}
}

@inproceedings{ance,
  title={Approximate nearest neighbor negative contrastive learning for dense text retrieval},
  author={Xiong, Lee and Xiong, Chenyan and Li, Ye and Tang, Kwok-Fung and Liu, Jialin and Bennett, Paul and Ahmed, Junaid and Overwijk, Arnold},
  booktitle={Proc. of ICLR},
  year={2021}
}

@inproceedings{retromae,
  title={RetroMAE: Pre-training retrieval-oriented language models via masked auto-encoder},
  author={Xiao, Shitao and Liu, Zheng and Shao, Yingxia and Cao, Zhao},
  booktitle={Proc. of EMNLP},
  year={2022}
}

@inproceedings{tasb,
  title={Balanced topic aware sampling for effective dense retriever: A reproducibility study},
  author={Wang, Shuai and Zuccon, Guido},
  booktitle={Proc. of SIGIR},
  year={2023}
}

@inproceedings{tctcolbert,
  title={In-batch negatives for knowledge distillation with tightly-coupled teachers for dense retrieval},
  author={Lin, Sheng-Chieh and Yang, Jheng-Hong and Lin, Jimmy},
  booktitle={RepL4NLP @ ACL},
  year={2021}
}

@article{e5,
  title={Text embeddings by weakly-supervised contrastive pre-training},
  author={Wang, Liang and Yang, Nan and Huang, Xiaolong and Jiao, Binxing and Yang, Linjun and Jiang, Daxin and Majumder, Rangan and Wei, Furu},
  journal={arXiv preprint arXiv:2212.03533},
  year={2022}
}

@article{splade_pp,
  title={Towards effective and efficient sparse neural information retrieval},
  author={Formal, Thibault and Lassance, Carlos and Piwowarski, Benjamin and Clinchant, St{\'e}phane},
  journal={TOIS},
  volume={42},
  number={5},
  year={2024}
}

@article{splade_v2,
  title={SPLADE v2: Sparse lexical and expansion model for information retrieval},
  author={Formal, Thibault and Lassance, Carlos and Piwowarski, Benjamin and Clinchant, St{\'e}phane},
  journal={arXiv preprint arXiv:2109.10086},
  year={2021}
}

@article{splade_v3,
  title={SPLADE-v3: New baselines for SPLADE},
  author={Lassance, Carlos and D{\'e}jean, Herv{\'e} and Formal, Thibault and Clinchant, St{\'e}phane},
  journal={arXiv preprint arXiv:2403.06789},
  year={2024}
}

@article{monot5,
  title={The expando-mono-duo design pattern for text ranking with pretrained sequence-to-sequence models},
  author={Pradeep, Ronak and Nogueira, Rodrigo and Lin, Jimmy},
  journal={arXiv preprint arXiv:2101.05667},
  year={2021}
}

@article{minilm,
  title={Minilm: Deep self-attention distillation for task-agnostic compression of pre-trained transformers},
  author={Wang, Wenhui and Wei, Furu and Dong, Li and Bao, Hangbo and Yang, Nan and Zhou, Ming},
  journal={NeurIPS},
  volume={33},
  year={2020}
}

@inproceedings{electra,
  title={Squeezing water from a stone: a bag of tricks for further improving cross-encoder effectiveness for reranking},
  author={Pradeep, Ronak and Liu, Yuqi and Zhang, Xinyu and Li, Yilin and Yates, Andrew and Lin, Jimmy},
  booktitle={Proc. of ECIR},
  year={2022}
}

@article{monobert,
  title={Multi-stage document ranking with BERT},
  author={Nogueira, Rodrigo and Yang, Wei and Cho, Kyunghyun and Lin, Jimmy},
  journal={arXiv preprint arXiv:1910.14424},
  year={2019}
}

@article{spearman_rho,
  title={The Proof and Measurement of Association between Two Things},
  author={Spearman, C},
  journal={AJP},
  volume={15},
  year={1904}
}

@article{kendall_tau,
  title={A new measure of rank correlation},
  author={Kendall, Maurice G},
  journal={Biometrika},
  volume={30},
  number={1-2},
  year={1938},
}

@inproceedings{tau_ap,
  title={A new rank correlation coefficient for information retrieval},
  author={Yilmaz, Emine and Aslam, Javed A and Robertson, Stephen},
  booktitle={Proc. of SIGIR},
  year={2008}
}

@inproceedings{faggioli_llm,
  title={Perspectives on large language models for relevance judgment},
  author={Faggioli, Guglielmo and Dietz, Laura and Clarke, Charles LA and Demartini, Gianluca and Hagen, Matthias and Hauff, Claudia and Kando, Noriko and Kanoulas, Evangelos and Potthast, Martin and Stein, Benno and others},
  booktitle={Proc. of ICTIR},
  year={2023}
}

@article{qwen,
  title={Qwen technical report},
  author={Bai, Jinze and Bai, Shuai and Chu, Yunfei and Cui, Zeyu and Dang, Kai and Deng, Xiaodong and Fan, Yang and Ge, Wenbin and Han, Yu and Huang, Fei and others},
  journal={arXiv preprint arXiv:2309.16609},
  year={2023}
}

@article{aya,
  title={Aya expanse: Combining research breakthroughs for a new multilingual frontier},
  author={Dang, John and Singh, Shivalika and D'souza, Daniel and Ahmadian, Arash and Salamanca, Alejandro and Smith, Madeline and Peppin, Aidan and Hong, Sungjin and Govindassamy, Manoj and Zhao, Terrence and others},
  journal={arXiv preprint arXiv:2412.04261},
  year={2024}
}

@article{email_audience,
  title={Configuring audiences: A case study of email communication},
  author={Zhang, Justine and Pennebaker, James and Dumais, Susan and Horvitz, Eric},
  journal={PACMHCI},
  volume={4},
  number={CSCW1},
  year={2020}
}

@article{gov_ai_sector,
  title={Artificial Intelligence Sector Deal},
  author={{Department for Science, Innovation and Technology} and {Department for Business and Trade} and {Office for Artificial Intelligence, Department for Digital, Culture, Media \& Sport} and {Department for Business, Energy \& Industrial Strategy}},
  journal={https://www.gov.uk/government/publications/artificial-intelligence-sector-deal},
  year={2019}
}

@article{lda,
  title={Inference of population structure using multilocus genotype data},
  author={Pritchard, Jonathan K and Stephens, Matthew and Donnelly, Peter},
  journal={Genetics},
  volume={155},
  number={2},
  year={2000}}

@inproceedings{kamps_foia,
  title={Enticing Local Governments to Produce FAIR Freedom of Information Act Dossiers},
  author={Marx, Maarten and Larooij, Maik and Perasedillo, Filipp and Kamps, Jaap},
  booktitle={Proc. of ECIR},
  year={2023}
}

@inproceedings{introducing_enron,
  title={Introducing the Enron corpus},
  author={Klimt, Bryan and Yang, Yiming},
  booktitle={Proc. of CEAS},
  year={2004}
}

@article{chuan_judges,
  title={Re-Rankers as Relevance Judges},
  author={Meng, Chuan and Liu, Jiqun and Aliannejadi, Mohammad and Mo, Fengran and Dalton, Jeff and de Rijke, Maarten},
  journal={arXiv preprint arXiv:2601.04455},
  year={2026}
}

@inproceedings{ecir_dc,
  title={Cascading Ranking Pipelines for Sensitivity-Aware Search},
  author={McKechnie, Jack},
  booktitle={Proc. of ECIR},
  year={2024}
}

\end{document}